\documentclass[preprint]{elsarticle}
\usepackage{graphics}
\usepackage{graphicx}
\usepackage{amsmath}
\usepackage{amssymb}
\usepackage{subfigure}
\usepackage{longtable}
\usepackage{multirow}
\usepackage{bm}
\usepackage{hyperref}

\newcommand{\ve}[1]{\bm{\mathrm{#1}}}

\newcommand{\w}{\omega}

\graphicspath{{../figures/}}

\begin{document}
\title{Graphics Processing Unit Acceleration of the Random Phase Approximation in the Projector Augmented Wave Method}
\author[rvt]{Jun Yan\corref{cor1}}
\ead{junyan@stanford.edu}
\cortext[cor1]{Corresponding author}
\author[rvt]{Lin Li}
\author[rvt]{Christopher O'Grady}
\address{SUNCAT Center for Interface Science and Catalysis,   \\
SLAC National Accelerator Laboratory \\ 2575 Sand Hill Road, Menlo Park, CA 94025, USA}
\date{\today}

\begin{abstract}

  The Random Phase Approximation (RPA) for correlation energy in the
  grid-based projector augmented wave (\textsc{gpaw}) code is
  accelerated by porting to the Graphics Processing Unit (GPU)
  architecture. The acceleration is achieved by grouping independent
  vectors/matrices and transforming the implementation from being
  memory bound to being computation/latency bound. With this approach, both the
  CPU and GPU implementations have been enhanced.  We tested the GPU
  implementation on a few representative systems: molecules (O$_2$), bulk
  solids (Li$_2$O and MoO$_3$) and molecules adsorbed on metal
  surfaces (N$_2$/Ru(0001) and CO/Ni(111)).  Improvements from $10\times$  to
  $40\times$ have been achieved
  (8-GPUs versus 8-CPUs) .  A realistic RPA calculation for CO/Ni(111)
  surface can be finished in 5.5 hours using 8 GPUs.  It
  is thus promising to employ non-self-consistent RPA for routine
  surface chemistry simulations.

\end{abstract}

\maketitle

\section{Introduction}

Density functional theory (DFT) has became one of the standard tools
for predicting the structural, energetic, mechanical, dielectric and
magnetic properties of condensed matter. In the field of computational
catalytic design, thousands of DFT calculations have been reported and
tabulated for different adsorption species on a variety of metal
surfaces with different facets \cite{Catapp}. The general trend of the
variations in catalytic activity from one catalyst to another is
reported to be well captured by DFT calculations \cite{Norskov_NC09}.
However, DFT does not yet achieve the chemical accuracy (1 kcal/mol or
0.0434 eV) needed to predict the energetics of many catalysts. This
level of accuracy is important because the reactivity depends
exponentially on adsorption and reaction energies. Furthermore, for
systems with strong correlation and localization, it is unclear if DFT
can qualitatively predict the correct catalytic trends.


Random Phase Approximation (RPA) has emerged as a promising approach
to improve the precision of total energy predictions in computational
chemistry and materials science
\cite{Xingguo_Rev12,Furche_Rev12,Gorling_Rev11}.  Instead of using the
local or semi-local exchange-correlation functionals of standard DFT
calculations, RPA accounts for dynamic electronic screening and is
fully non-local. It has been shown to systematically improve lattice
constants \cite{Kresse_B10}, atomization and cohesive energies
\cite{Kresse_B08}, adsorption sites and energies
\cite{Scheffler_B09,Kresse_NM10,Sergey_B12}, reaction barriers
\cite{Xingguo_Rev12}, and structural phase
transitions\cite{Perdew_B12} for a wide range of systems that have
ionic, covalent and/or van der Waals interactions
\cite{Kresse_L09,Olsen_L11,Dobson_Rev12,Galli_L09}. Similar to many other
beyond-DFT calculations, the RPA method is exceptionally computational
demanding since it involves summations over large basis sets and
hundreds or thousands of unoccupied orbitals, both of which are
truncated above a certain energy.

The most demanding part of an RPA calculation is to compute the
non-interacting response function. There are a few methods proposed in
the literature to speed up the calculation of the response function.
These methods focus on reducing or eliminating the number of
unoccupied orbitals, and have been applied to compute the
self-energies in GW calculations \cite{Gonze_B08,Reining_B10}.
However, GW self-energies are known not to depend strongly on the
approximations made to the imaginary part of the response function. It
is unknown how such approximations would affect RPA calculations. A
more rigorous algorithm using many-body perturbation theory can
completely eliminate the unoccupied orbitals \cite{Baroni_B10,
  Louie_B10, Galli_B12}, however the algorithm requires solving a complex eigenvalue equation 
  (either by diagonalizing or using a Lanczos scheme), and the number of basis functions used cannot be
reduced. Recently, Bruneval proposed a range-separated approach:
preserving the RPA long-range non-local correlation (which is the
essence of the RPA method and is fast to converge in reciprocal space)
and using a local functional to replace the short range
correlation\cite{Bruneval_L12}.  Such an approach significantly speeds
up the convergence with respect to the number of states and has been
applied to a silicon vacancy in a 216 atom supercell.

Graphics processing units (GPUs) are becoming attractive platforms for
high performance scientific computing. A GPU contains hundreds of
cores, with low price to performance ratio and a relatively low energy
consumption per core compared to traditional CPUs.
They are most advantageous for parallelizable algorithms  requiring a large number
of numerical operations per memory fetch (``numerically-intensive'').
An increasing number of scientific applications have been
ported to GPUs \cite{GPU_book}. In particular, for electronic structure
calculations, GPUs have been utilized for many quantum chemistry
and DFT codes such as \textsc{gpaw} \cite{gpaw_gpu}, \textsc{vasp}
\cite{vasp_gpu1, vasp_gpu2}, \textsc{quantum espresso} \cite{QE_gpu},
\textsc{terachem} \cite{terachem}, \textsc{bigdft} \cite{bigdft_gpu},
\textsc{petot} \cite{petot_gpu} and \textsc{octopus}
\cite{octopus_gpu}. The speed up of these DFT-based electronic
structure codes is generally less than $15\times$ due to the complexity and communication bottlenecks in algorithms such as FFT, subspace diagonalization, minimization and orbital orthorgonalization. 
These problems, however, do not apply to beyond DFT methods based on
the linear density response function such as the RPA correlation energy.
The evaluation of the linear density response matrix $\chi^0_{\ve G \ve G^{\prime}}(\ve
q, \w)$ is in general numerically-intensive. The single particle
transitions that are required to calculate the response matrix are
completely independent of each other.  This allows for 
straightforward MPI parallelization, with no MPI communication required
during the calculation. It is thus natural to port the RPA method to
GPUs.

This paper describes the GPU porting of the RPA method (specifically,
the evaluation of the linear density response function) as well as
the performance on a few representative molecules, bulk solids and
surfaces. The rest of the paper is organized as follows.  In Section
\ref{sec:rpa_method} the theory and algorithm of RPA is briefly
reviewed, focusing on the CPU implementation. Section \ref{sec:gpu}
presents a simpler ``direct'' GPU porting, which is achieved with
minimal changes to the code structure, followed by a ``multi-$u$''
technique that enhances both the GPU and CPU implementations.  The
performance of the GPU implementation for a few representative systems
is discussed in Section \ref{sec:performance} and finally conclusions
are given in Section \ref{sec:conclusion}.

\section{The RPA Method and the CPU  Implementation}
\label{sec:rpa_method}
The RPA scheme for obtaining the total energy consists of two parts:
exact exchange energy and correlation energy using RPA.  Both can be
derived from the adiabatic connection fluctuation-dissipation theory
(ACFDT)\cite{Gonze_A03}.
Here we only briefly review the portion of RPA theory that is
relevant to the GPU porting of the code, and focus instead on the
actual implementation.
The theory on the exact exchange energy and its porting to the GPU
platform can be found in Ref. \cite{exx_gpu}.

\subsection{Correlation energy using random phase approximation}
According to ACFDT, the RPA correlation energy
$E_{\mathrm{rpa}}^{\mathrm{c}}$ can be formulated as
\begin{equation}\label{Eq:rpa}
 E_{\mathrm{rpa}}^{\mathrm{c}} = \int_0^{\infty}\frac{d\omega}{2\pi}
 \mathrm{Tr}\{\mathrm{ln} [1- v \chi^0(i \w) ]
 + v \chi^0(i \w) \},
\end{equation}
where $v$ is the Coulomb interaction kernel and $\chi^0$ is the
non-interacting response function. $\chi^0$ is a fundamental quantity
in many beyond-DFT methods and is described in the following section.

In the \textsc{gpaw} \cite{GPAW_B05, GPAW_10} RPA implementation
\cite{Jun_response}, $v$ and $\chi^0$ are represented using a plane
wave basis set and become $v_{\ve G}(\ve q)$ and $\chi^0_{\ve G \ve
  G^{\prime}}(\ve q, i\w)$, respectively. $\ve q$ is a wave vector
within the Brillouin zone (BZ) and $\ve G$ is a reciprocal space lattice vector,
the size of which is defined by a cutoff energy $E_{\mathrm{cut}}$.
The RPA correlation energy under the plane wave representation becomes
\begin{equation}\label{Eq:rpa_c}
 E_{\mathrm{rpa}}^{\mathrm{c}} = \int_0^{\infty} \frac{d\omega}{2\pi} 
 \int_{\mathrm{BZ}} d\ve q \mathrm{Tr} \{ \mathrm{ln}[1-v_{\ve G}(\ve q)\chi^0_{\ve G \ve G^{\prime}}(\ve q, i\omega)] + v_{\ve G}(\ve q)\chi^0_{\ve G \ve G^{\prime}}(\ve q, i\omega)  \}
\end{equation}
The frequency integration over $\omega$ is carried out using 16
Gauss-Legendre points following the procedure from Ref.
\cite{Olsen_L11}. The integration over the BZ is discretized using
Monkhorst-pack $k$-points. By exploiting the $\ve q$-mesh symmetry,
the integration is reduced to a summation over the irreducible BZ (IBZ) :
$\int_{\mathrm{BZ}} \rightarrow \sum_{\mathrm{IBZ}} w_{\ve q}$ where
$w_{\ve q}$ is the weight for a specific $\ve q$ vector.

\subsection{Density response function in the projector-augmented wave method}
The non-interacting response function is evaluated in a plane wave basis
using the ``sum over states" approach written as
\begin{equation}\label{Eq:chi0}
  \chi^0_{\ve G \ve G^{\prime}}(\ve q, i \w) = \frac{2}{\Omega}\sum_{\ve k}^{\mathrm{BZ}} \sum_{n n^{\prime}}\frac{f_{n\ve k} - f_{n^{\prime} \ve k+\ve q}}{i \w + \epsilon_{n\ve k} - \epsilon_{n^{\prime} \ve k + \ve q}}  
n_{n\ve k, n^{\prime}\ve k + \ve q} (\ve G)
n^{\ast}_{n\ve k, n^{\prime}\ve k + \ve q} (\ve G^{\prime})
\end{equation}
where
\begin{equation}\label{Eq:n_G}
 n_{n\ve k, n^{\prime}\ve k + \ve q} (\ve G)\equiv \left< \psi_{n\ve k} \right | e^{-i(\ve q +\ve G)\cdot \ve r} \left | \psi_{n^{\prime} \ve k + \ve q} \right> 
\end{equation} 
is the charge density matrix and $\Omega$ is the volume of the unit cell.  The occupation $f_{n\ve k}$, Kohn-Sham
(KS) eigenvalue $\epsilon_{n\ve k}$ and eigenstate $\psi_{n\ve k}$ for
band $n$ at wave vector $\ve k$ are extracted from a DFT calculation.
The spin index $\sigma$ is implicitly contained in the above formula.

In the PAW formalism \cite{Blochl_B94}, a true all-electron KS wave
function $\psi_{n\ve k}$ is obtained by a linear transformation $\mathcal{T}$ from a
smooth pseudo-wave function $\tilde{\psi}_{n\ve k}$ via $\psi_{n\ve
  k}= \mathcal{T} \tilde{\psi}_{n\ve k}$. The transformation operator
is chosen in a manner such that the all electron wave function
$\psi_{n\ve k}$ is the sum of the pseudo one $ \tilde{\psi}_{n\ve k}$
plus a contribution centered around each atom written as
\begin{equation}\label{Eq:transformation}
\psi_{n\ve k} (\ve r)= \tilde{\psi}_{n\ve k} (\ve r)+ \sum_{a, i} \langle \tilde{p_i^a} | \tilde{\psi}_{n\ve k} \rangle [\phi_i^a(\ve r - \ve R_a) - \tilde{\phi}_i^a(\ve r - \ve R_a)]
\end{equation}
The pseudo-wave function $\tilde{\psi}_{n\ve k}$ matches the
all-electron one $\psi_{n\ve k}$ outside the augmentation spheres
centered on each atom $a$ at position $\ve R_a$. Their differences
inside the augmentation spheres are expanded on atom-centered
all-electron partial waves $\phi_i^a$ and the smooth counterparts
$\tilde{\phi}_i^a$. The expansion coefficient is given by $\langle
\tilde{p_i^a} | \tilde{\psi}_{n\ve k} \rangle$, where $ \tilde{p_i^a}$
is a dual basis to the pseudo-partial wave and is called a projector
function. In the practical \textsc{gpaw} implementation, the
pseudo-wave function can be discretized using plane waves,
three-dimensional uniform real space grids, or a localized atomic
orbital basis. The atomic centered quantities such as $\tilde{p_i^a}$,
$\phi_i^a$ and $\tilde{\phi}_i^a$ are expanded using a one dimensional
logarithmic grid, which becomes denser closer to the atomic cores.

Substituting Eq. (\ref{Eq:transformation}) into Eq. (\ref{Eq:n_G}),
the charge density matrix becomes
\begin{equation}\label{Eq:n_G_2}
 n_{n\ve k, n^{\prime}\ve k + \ve q} (\ve G) =  
 \tilde{n}_{n\ve k, n^{\prime}\ve k + \ve q} (\ve G) 
 + \sum_{a, ij} \langle \tilde{\psi}_{n\ve k} | \tilde{p}^a_i \rangle 
 \langle \tilde{p}^a_j  | \tilde{\psi}_{n^{\prime}\ve k+\ve q}  \rangle 
 Q_{ij}^a(\ve q + \ve G)
\end{equation}
with definitions
\begin{equation}
 \tilde{n}_{n\ve k, n^{\prime}\ve k + \ve q} (\ve G)  \equiv \langle\tilde{\psi}_{n\ve k} | e^{-i(\ve q +\ve G)\cdot \ve r} | \tilde{\psi}_{n^{\prime} \ve k + \ve q} \rangle 
\end{equation}
\begin{equation}
Q_{ij}^a(\ve q + \ve G) \equiv \langle \phi_i^a | e^{-i (\ve q+\ve G) \cdot \ve r} | \phi_j^a \rangle - 
 \langle \tilde{\phi}_i^a | e^{-i (\ve q+\ve G) \cdot \ve r} | \tilde{\phi}_j^a \rangle. 
\end{equation}
In the above equations, $\ve k$ and $\ve k + \ve q$  are wave
vectors within the BZ; however in a general DFT calculation the
eigenvalues and eigenstates are computed only for $k$-points in the IBZ.
As a result, a mapping of the $k$-point indices from BZ to IBZ is
required:
\begin{equation}\label{Eq:kmap}
 \ve k = T_1 \ve k_1^{\mathrm{IBZ}}, \ve k + \ve q = T_2 \ve k_2^{\mathrm{IBZ}},
\end{equation}
where $T_1$ and $T_2$ are the transformation operators for $\ve k$ and
$\ve k + \ve q$, respectively.  Correspondingly, a transformation of
the KS eigenvectors from IBZ to BZ using crystal symmetry operations
is needed and will be described in the following.

For the RPA implementation, the pseudo-wave function is expanded using
a plane wave basis. This choice is motivated by the fact that a
typical RPA correlation energy calculation requires hundreds to
thousands of eigenstates, which can be efficiently computed by direct
diagonalization of the KS hamiltonian in a plane wave basis using a
previously converged self-consistent DFT calculation. Given plane wave
coefficients $C_{n\ve k_1^{\mathrm{IBZ}}}(\ve Q)$ and $C_{n\ve
  k_2^{\mathrm{IBZ}}}(\ve Q^{\prime})$, where $\ve Q$ and $\ve
Q^{\prime}$ are reciprocal lattice vectors, the pseudo-wave function
in the IBZ is obtained through
\begin{equation}\label{Eq:get_wfs}
\tilde{\psi}_{n\ve k_1^{\mathrm{IBZ}}}(\ve r) = \sum_{\ve Q} C_{n\ve k_1^{\mathrm{IBZ}}}(\ve Q) e^{i (\ve k_1^{\mathrm{IBZ}}+\ve Q) \cdot \ve r}, 
\tilde{\psi}_{n^{\prime}\ve k_2^{\mathrm{IBZ}}}(\ve r) = \sum_{\ve Q^{\prime}} C_{n^{\prime}\ve k_2^{\mathrm{IBZ}}}(\ve Q^{\prime}) e^{i (\ve k_2^{\mathrm{IBZ}}+ \ve Q^{\prime}) \cdot \ve r}, 
\end{equation}
which can be performed efficiently using a Fast Fourier Transform (FFT). 
The transformation of the pseudo-wave function from IBZ to BZ is
\begin{equation}\label{Eq:trans_wfs}
  \tilde{\psi}_{n \ve k}(\ve r)  = \tilde{\psi}_{n\ve k_1^{\mathrm{IBZ}}}( T_1^{-1}\ve r), 
  \tilde{\psi}_{n^{\prime} \ve k+\ve q}(\ve r)  = \tilde{\psi}_{n^{\prime}\ve k_2^{\mathrm{IBZ}}}( T_2^{-1}\ve r),
\end{equation}
Both crystal symmetries and time-reversal symmetry are taken into
account in the above transformation.

After obtaining the $\tilde{\psi}_{n \ve k}(\ve r)$ and
$\tilde{\psi}_{n^{\prime} \ve k+\ve q}(\ve r)$ on the uniform 3D grid,
the pseudo-density matrix $\tilde{n}_{n\ve k, n^{\prime}\ve k + \ve q}
(\ve G)$ is obtained using an FFT $\mathcal{F}$ through
\begin{equation}\label{Eq:fft}
\tilde{n}_{n\ve k, n^{\prime}\ve k + \ve q} (\ve G) = \mathcal{F}[\tilde{\psi}^{\ast}_{n\ve k}(\ve r)
\tilde{\psi}_{n^{\prime}\ve k + \ve q} (\ve r) e^{-i\ve q \cdot \ve r}]. 
\end{equation}
Compared to directly integrating the wave functions for each $\ve G$
on the 3D grid, the use of an FFT achieves a speed up $>100\times$.

For the augmentation part in Eq. \ref{Eq:n_G_2}, the $\langle
\tilde{p}^a_i | \tilde{\psi}_{n\ve k_1^{\mathrm{IBZ}}} \rangle $ and
$\langle \tilde{p}^a_j | \tilde{\psi}_{n^{\prime}\ve
  k_2^{\mathrm{IBZ}}} \rangle $ are calculated and saved during DFT
calculations for $\ve k$-points that are inside the IBZ. For $\ve
k$-points outside the IBZ, symmetry operations are applied to the
projector function $\tilde{p}_i^a$ and $\tilde{p}_j^a$ to obtain
$\langle \tilde{p}^a_i | \tilde{\psi}_{n\ve k} \rangle $ and $\langle
\tilde{p}^a_j | \tilde{\psi}_{n^{\prime}\ve k + \ve q} \rangle $,
respectively. The $Q_{ij}^a(\ve q + \ve G)$ is calculated on a 1D
logarithmic grid by expanding the $e^{-i\ve (\ve q + \ve G) \cdot \ve
  r}$ using a spherical harmonic basis\cite{Jun_response}. Since it
requires only a single shot calculation, it is performed at the
initialization step of an RPA calculation on the CPU and then copied
to the GPU.

For the case $\ve q=0$ and $\ve G=0$, the charge density matrix in
Eq. (\ref{Eq:n_G}) becomes $\langle \psi_{n\ve k} |
\psi_{n^{\prime}\ve k} \rangle = \delta_{nn^{\prime}}$; on the other
hand, the coulomb kernel $v_{\ve G}(\ve q) = 4\pi/|\ve q + \ve G|^2$
becomes divergent. To cure the divergence of the coulomb kernel, a
perturbative approach is used by taking the $\ve q \rightarrow 0$
limit and the charge density matrix is calculated using
\begin{eqnarray}\label{Eq:n_q0}
n_{\ve q\rightarrow 0}(\ve G=0) &=& \langle \psi_{n\ve k} | e^{-i\ve q\cdot \ve r} | \psi_{n^{\prime}\ve k+\ve q} \rangle_{\ve q\rightarrow 0} \\
&=& \frac{-i \ve q \cdot \langle \psi_{n\ve k} | \nabla | \psi_{n^{\prime}\ve k} \rangle }  {\epsilon_{n^{\prime} \ve k} - \epsilon_{n\ve k}}
\end{eqnarray}
In the above derivation,  $\psi_{n^{\prime}\ve k+\ve q} $ at $\ve q
\rightarrow 0$ is expanded using  $k\cdot p$ second order
perturbation theory \cite{Jun_response}. In the PAW method, the matrix
element $\langle \psi_{n\ve k} | \nabla | \psi_{n^{\prime}\ve k}
\rangle$ is given by
\begin{equation}\label{Eq:nabla}
\langle \psi_{n\ve k} | \nabla | \psi_{n^{\prime}\ve k} \rangle = 
\langle \tilde{\psi}_{n\ve k} | \nabla | \tilde{\psi}_{n^{\prime}\ve k} \rangle 
+ \sum_{a, ij} \langle \tilde{\psi}_{n\ve k} | \tilde{p}^a_i \rangle 
 \langle \tilde{p}^a_j  | \tilde{\psi}_{n^{\prime}\ve k}  \rangle 
 Q_{ij}^a(\ve G=0)_{\ve q\rightarrow 0}
\end{equation}
with  
\begin{equation}\label{Eq:Q_q0}
 Q_{ij}^a(\ve G=0)_{\ve q\rightarrow 0} =  \langle \phi_i^a | \nabla | \phi_j^a \rangle - 
 \langle \tilde{\phi}_i^a | \nabla | \tilde{\phi}_j^a \rangle 
\end{equation}
The pseudo-wave function part $\langle \tilde{\psi}_{n\ve k} | \nabla
| \tilde{\psi}_{n^{\prime}\ve k} \rangle$ is calculated using a finite
difference approximation for the nabla operator, taking into account 6
neighboring points (in total 13 grid points) in each direction. The
$Q_{ij}^a(\ve G=0)_{\ve q\rightarrow 0}$ is calculated by expanding
the partial waves ($\phi_i^a$, $ \tilde{\phi}_i^a$) on real spherical
harmonics and applying the nabla operator on the radial and angular
part of the expansion coefficients. For a detailed derivation of the
$Q_{ij}^a(\ve G=0)_{\ve q\rightarrow 0}$ limit on a 1D logarithmic
grid, refer to Ref.  \cite{Jun_response}.

\subsection{The CPU implementation flow}

\begin{table}[h]
\caption{\label{Tb:flow} RPA algorithm}
\vspace{0.2cm}
\begin{tabular*}{1.0\linewidth}{r p{0.9\linewidth}  }
\hline
1) & Initialization, including MPI initialization and distribution;  \\
2) & Read plane wave coefficients $C_{n\ve k_1^{\mathrm{IBZ}}}(\ve Q)$ and 
$C_{n^{\prime}\ve k_2^{\mathrm{IBZ}}}(\ve Q^{\prime})$, where $\ve k_1^{\mathrm{IBZ}}$ and $\ve k_2^{\mathrm{IBZ}}$
are the corresponding IBZ $k$-points for $\ve k$ and $\ve k+ \ve q$, respectively, following Eq. (\ref{Eq:kmap});  \\
3) & Calculate (using FFT) the pseudo-wave function $\tilde{\psi}_{n\ve k_1^{\mathrm{IBZ}}}(\ve r)$ and 
$\tilde{\psi}_{n^{\prime}\ve k_2^{\mathrm{IBZ}}}(\ve r)$ from the coefficients $C_{n\ve k_1^{\mathrm{IBZ}}}(\ve Q)$ and 
$C_{n^{\prime}\ve k_2^{\mathrm{IBZ}}}(\ve Q^{\prime})$, respectively, following Eq. (\ref{Eq:get_wfs});   \\
4) & Transform the pseudo-wave function from IBZ to BZ and obtain $\tilde{\psi}_{n\ve k}(\ve r)$ and $\tilde{\psi}_{n^{\prime}\ve k+\ve q}(\ve r)$ according to Eq. (\ref{Eq:trans_wfs}); \\
5) & Calculate pseudo density matrix using FFT according to Eq. (\ref{Eq:fft}); \\
6) & Map the FFT result on the FFT grid to a reduced grid $\ve G$, the size of which is defined  by the cutoff energy of the response function,  to get $\tilde{n}(\ve G)$; \\
7) &Read $\langle \tilde{p}_i^a | \tilde{\psi}_{n\ve k_1^{\mathrm{IBZ}}} \rangle$ and $\langle \tilde{p}_j^a  | \tilde{\psi}_{n^{\prime}\ve k_2^{\mathrm{IBZ}}}  \rangle$, transform them from IBZ to BZ to obtain $P(a, i)\equiv \langle \tilde{p}_i^a | \tilde{\psi}_{n\ve k} \rangle $ and $P(a, j) \equiv \langle \tilde{p}_j^a | \tilde{\psi}_{n\ve k+\ve q} \rangle$, respectively; \\
8) & Perform  $P(a, p) \equiv P^{\ast}(a, i) \otimes P(a, j)$, where $p \equiv {\{ij\}}$ is a combined index of ${ij}$;  \\
9) & Perform $\tilde{n}(\ve G) +=\sum_{ap} P(a, p) Q(a, p, \ve G)$. Step 8) and 9) follow Eq. (\ref{Eq:n_G_2}); \\
10) & If $\ve q=0$, calculate and replace $n(\ve G=0)$ using Eq. (\ref{Eq:n_q0}) - (\ref{Eq:Q_q0}); otherwise skip this step; \\
11) & Perform $\chi^0(i\w, \ve G, \ve G^{\prime}) += A(i\omega)n(\ve G)n^{\ast}(\ve G^{\prime})$ following Eq. (\ref{Eq:zher}) - (\ref{Eq:A_iw}); \\
12) & Steps 2 - 11 are looped over $n$, $n^{\prime}$, $\ve k$ and $s$ (spin, not explicitly written) indices until the calculation of $\chi^0$ at a particular $\ve q$ is finished; \\
13) & Compute the contribution to $E^c_{\mathrm{rpa}}$ at the particular $\ve q$ according to Eq. (\ref{Eq:rpa_c}); \\
14) & Steps 12 - 13 are looped over $\ve q$ until $E^c_{\mathrm{rpa}}$ is finished. \\
\hline
\end{tabular*}
\end{table}

The response function $\chi^0_{\ve G \ve G^{\prime}}(\ve q, i \w) $ in
Eq. (\ref{Eq:chi0}) is implemented as a double precision matrix of
size $(nG)^2 \times nq \times n\omega$, where $nG, nq, n\omega$
corresponds to the number of plane waves, q-points and frequency
points, respectively.  The sizes of $nG$ and $nq$ are system
dependent, and are typically a few hundreds to thousands, and a few
tens to hundreds, respectively. $n\omega$ corresponds to the 16 Gauss-Legendre 
points.  Considering that a typical memory of 2-3 Gigabytes
is available per core, we choose to loop over q-points and store the
matrix $\chi^0_{\ve G \ve G^{\prime}}(i \w) $ in memory during
computations.

In order to obtain $\chi^0_{\ve G \ve G^{\prime}}(i \w)$ at a given
$\ve q$, we need to compute the charge density matrix $n_{n\ve k,
  n^{\prime}\ve k + \ve q} (\ve G)$ in Eq. (\ref{Eq:n_G}). The number
of bands (index $n$ and $n^{\prime}$) and number of $k$-points (index
$k$) are generally too large, so the entire density matrix can not
reside in memory for RPA calculations. As a result, the computation of
$\chi^0_{\ve G \ve G^{\prime}}(i \w)$ is achieved by looping and
summing over $n$, $n^{\prime}$, $k$ and $s$ (spin, implicitly
included) indices, and calculating the charge density matrix, which is
in fact a vector $n(\ve G)$ of length $nG$, within each loop according
to
\begin{equation}\label{Eq:zher}
 \chi^0_{\ve G \ve G^{\prime}}(i \w) = \sum_{\ve k, nn^{\prime}} A(i\omega) n(\ve G) n^{\ast}(\ve G^{\prime}),
\end{equation}
where  $A(i\omega)$ is a vector defined as (given $n$, $n^{\prime}$, $\ve k$ and $\ve q$)
\begin{equation}\label{Eq:A_iw}
 A(i\omega)  \equiv \frac{2}{\Omega} \times \frac{f_{n\ve k} - f_{n^{\prime} \ve k+\ve q}}{i \w + \epsilon_{n\ve k} - \epsilon_{n^{\prime} \ve k + \ve q}}  
\end{equation} 

Table \ref{Tb:flow} shows the algorithm used to compute an RPA
correlation energy. In order to make the size of the matrix clearer,
we change the notation for the matrices in Table \ref{Tb:flow} so that
the values inside the parentheses correspond to the size of the
matrix. For instance, $C_{n\ve k_1^{\mathrm{IBZ}}}(\ve Q)$ is a vector
of length $NQ$. It is, however, slightly different for PAW related
functions because the number of projector functions $P_i^a$ is
different for each atom. For example, $P(a, i)$ represents a list of
atoms of length $Na$ and for each atom, a vector of size $Ni$; while
$Q(a, p, \ve G)$ represents a list of atoms of length $Na$ and for
each atom, a matrix of size $Np * NG$.

Steps 2 - 11 take more than 99.9\% of the total computing time and we
focus on this part. Step 2 performance is determined by the I/O speed of reading
orbitals from a previously saved DFT calculation. Step 3 is separated
into two parts: first mapping the coefficient to the FFT grid and then
performing the FFT. The transformation in step 4 corresponds to a
mapping from one $\ve r$ grid to another: $\ve r^{\prime} =
T^{-1}\ve r$. Step 5 contains two parts: evaluating
$\tilde{\psi}^{\ast}_{n\ve k}(\ve r) \tilde{\psi}_{n^{\prime}\ve k +
  \ve q} (\ve r) e^{-i\ve q \cdot \ve r}$ on the 3D uniform real space
grid and performing a 3D FFT. Step 6 maps the result on the 3D FFT
grid to a reduced $\ve G$ 1D grid. Steps 7 - 9 calculate the PAW
corrections to the response function, with steps 8 and 9 corresponding
to an outer and inner product of two functions for each atom,
respectively, and finally a sum over atoms. Step 10 calculates the
optical limit correction to the response function at $\ve G=0$. 
Step 11 performs an outer product of two vectors.  The \textsc{blas}
library is exploited in step 8 (\textsc{gemm}), step 9 (\textsc{gemv}),
and step 11 (\textsc{zher}), while the \textsc{fftw} library is used
for FFTs in steps 3 and 5.

\section{Porting the RPA code to GPUs}\label{sec:gpu}
Here we present two separate steps in the GPU porting process.  A
``direct'' approach which makes minimal changes to the code, and a
``multi-$u$'' approach that uses higher performance
computation/latency bound algorithms instead of memory bound
algorithms, but requires more changes to the code structure.

\subsection{Direct GPU porting}
Direct GPU porting includes: replacing all the \textsc{mkl}
\textsc{blas} function calls with \textsc{cublas} and \textsc{fftw}
(version 3) with \textsc{cufft} (CUDA 5.0), as well as implementing a
few cuda kernels such as wave function transformation, index mapping and
and PAW projection evaluation. Double and double complex precision is used for
both the CPU and GPU code.  For each step, the GPU code maintains the
same data structure and flow as the CPU code, except for step 10.
The optical limit calculation for step 10 is performed using a finite
difference method (the so called ``stencil" method) for the derivative
operator in Eq. (\ref{Eq:nabla}). Instead of writing a cumbersome
stencil kernel using \textsc{cuda}, we reformulated the problem using
an FFT. Given that the pseudo-wave function can be expanded in plane
waves according to Eq. (\ref{Eq:get_wfs}), one can write
\begin{equation}
\nabla \tilde{\psi}_{n^{\prime}\ve k}(\ve r) = \nabla \left[\sum_{\ve Q} C_{n^{\prime}\ve k}(\ve Q) e^{i (\ve k + \ve Q) \cdot \ve r}\right] = \sum_{\ve Q} \left[ i (\ve k + \ve Q) \cdot C_{n^{\prime}\ve k}(\ve Q) \right] e^{i (\ve k + \ve Q) \cdot \ve r}, 
\end{equation}
which is efficiently evaluated using an FFT. Note that in the case
where $\ve k$ does not belong to the IBZ, a pseudo-wave function
transformation has to be performed first according to Eq.
(\ref{Eq:trans_wfs}).

\begin{table}[t]
\caption{\label{Tb:nmultix1} The timing (in units of seconds) for 1-CPU, 8-CPUs, 8-GPUs, as well as the 8-GPUs vs. 8-CPUs speed up (last column) for steps 2 - 11 in Table \ref{Tb:flow} for the test system N$_2$/Ru(0001) surface, modeled with 4 layers of Ru in a $\sqrt{3} \times \sqrt{3}$ unit cell. The timing information comes from a summation of 1 $k$-point (per  core), 5 occupied and 1486 unoccupied bands with an energy cutoff of 150 eV. The bottom of the table summarizes the total timing for both the optical limit ($\ve q\rightarrow 0$), and the $\ve q \neq 0$ calculations. Note that in order to measure the GPU time for each step $cudaDeviceSynchronize$ was used, while the total timing was obtained in the asynchronous mode. As a result, the ``Total" time is smaller than the sum of the individual times.  The CPU is an Intel Xeon X5650 and the GPU model is the ``C2075".
  }
\vspace{0.3cm}
\begin{minipage}{1.0\linewidth}
\begin{tabular*}{1.0\linewidth}{rlccccc}
\hline\hline
 No. & Function  & 1-CPU  & 8-CPUs   & 8-GPUs\footnote{1-CPU calculates 1 $k$-point while 8-CPUs/8-GPUs calculate 8 $k$-points.}  & Speed up \\
     &              &(seconds)            &  (seconds) & (seconds) &     (8-GPUs / 8-CPUs) \\
\hline   
          2) & read\_coef &      - & -    &    2.6 &   - \\
            3) & get\_wfs &     26.7 & 47.1 &    4.0 &  11.6$\times$ \\
      4) & transform\_wfs &    3.0 &  8.3 &    0.6 &  12.7$\times$ \\
                 5) & fft &     18.9 & 30.1 &    2.7 &  11.1$\times$ \\
                6) & mapG &   0.2 &    0.4 &    0.2 &   2.2$\times$ \\
          7) & paw\_P\_ai &   6.0 &  7.3 &    2.6 &   2.8$\times$ \\
          8) & paw\_P\_ap &   4.1 & 9.9 &    6.5 &   1.5$\times$ \\
            9) & paw\_add &   91.4 & 239.1 &   42.8 &   5.6$\times$ \\
     10) & optical\_limit &    197.9 & 267.8 &   50.9 &   5.3$\times$ \\
               11) & zher &  552.0 & 1193.2 &   89.6 &  13.3$\times$ \\
\hline
& Total,  $\ve q \rightarrow 0$ & 911.3  & 1816.7 &   188.7 &   9.6$\times$ \\
& Total, $\ve q \neq 0$ & 665.9 & 1545.7 &   123.2 &  12.5$\times$   \\ 
\hline \hline
\end{tabular*}
\end{minipage}
\end{table}

Table \ref{Tb:nmultix1} shows the timing results for the direct GPU
porting.  MPI initialization and distribution of the data was only
performed at step 1 and there is no MPI communication throughout steps
2 - 11. As a result, the timing for the 1-CPU and 8-CPUs cases are
expected to be the same.  However, as shown in Table
\ref{Tb:nmultix1}, the 8-CPUs case shows a $2$-$3\times$ slower performance
because the different cores within 1-node compete for CPU memory
bandwidth. In contrast, each GPU has its own dedicated memory, so
8-GPUs have the same timing as 1-GPU (results not shown).  In reality,
the RPA calculations will be executed on multiple nodes. As a result,
the 8-GPUs / 8-CPUs comparison is more relevant
than the 1-GPU / 1-CPU comparison. Also note that the CPU results were obtained with the number of
\textsc{mkl} threads set to 1. This is because threading is not
implemented for \textsc{zher} in our current \textsc{mkl} library (version 10.3), 
and we use MPI to parallelize over cores. 

According to Table \ref{Tb:nmultix1}, the average speed up for the
pseudo-wave function portion (steps 3, 4, 5, 6, 10, 11) is around
$10\times$, while for the PAW part (steps 7 - 9) it is around
$4\times$.  The most timing consuming part, step 11, which takes more
than 60\% of total simulation time, gains a speed up of only
$11.4\times$.  This is because the \textsc{cublas} \textsc{zher}
routine, which performs an outer product of a vector with length $n$
and adds that to a matrix of size $(n, n)$, is a memory bound operation.
Such an argument applies to the other \textsc{cublas} routines as
well. The extremely poor performance for the PAW portion arises
because the PAW contribution to the response function from each atom
has a matrix size which is typically 5-50 (number of projector
functions per atom) which is too small. In this case, the driver
overhead for the cuda kernels exceeds the execution time.

\subsection{Enhancing the GPU implementation}
The GPU timing in Table \ref{Tb:nmultix1} is dominated by steps 9 -
11, which use double-complex \textsc{cublas} (\textsc{gemv} and
\textsc{zher} routine) and the \textsc{cufft} library. Both
\textsc{gemv} and \textsc{zher} are memory bound routines. Since the
$n(\ve G)$ in Eq. (\ref{Eq:zher}) for each loop is completely
independent from the other loops, we can group different $n(\ve G)$
together such that
\begin{equation}\label{Eq:zher_multi}
 \chi^0_{\ve G \ve G^{\prime}}(i \w) = \sum_{\ve k, n, u\subset n^{\prime}} A(u, i\omega) n(u, \ve G) n^{\ast}(u, \ve G^{\prime}),
\end{equation}
where $u$ is a subset of index $n^{\prime}$, and $n(u, \ve G)$ is a
matrix, with each column representing a vector $n(\ve G)$ at a
particular $n^{\prime}$. We call this the ``multi-$u$'' approach\footnote{A similar approach is used in the GPU implementation of the \textsc{octopus} package\cite{octopus_gpu}, where the Kohn-Sham eigenstates are grouped together for time propagation of the Schr\"odinger equation.}. As a
result, a \textsc{blas} level 2 \textsc{zher} problem is transformed into a \textsc{blas}
level 3 \textsc{zherk} problem.

\begin{table}[t]
\caption{\label{Tb:multix} The timing (in units of seconds) and speed up (with respect to the $Nu=1$ time) for the ``multi-$u$'' approach, with $Nu=1$, $Nu=50$ and $Nu=250$ for the same steps and test system presented in Table \ref{Tb:nmultix1}. 8-GPUs are used throughout. 
}

\vspace{0.2cm}
\begin{center}
\begin{tabular}{rlccccc}
\hline\hline
 No. & Function  & $Nu=1$           &\multicolumn{2}{c}{$Nu=50$}          & \multicolumn{2}{c}{$Nu=250$}   \\
        &              &seconds  &  seconds & speed up  & seconds & speed up \\
\hline   
          2) & read\_coef &     2.5 &    2.6 &   1.0$\times$ &     3.1 &   0.8$\times$ \\
            3) & get\_wfs &     3.8 &    2.6 &   1.5$\times$ &     2.6 &   1.4$\times$ \\
      4) & transform\_wfs &     0.6 &    0.4 &   1.6$\times$ &     0.4 &   1.6$\times$ \\
                 5) & fft &     2.7 &    2.0 &   1.3$\times$ &     2.0 &   1.3$\times$ \\
                6) & mapG &     0.2 &    0.01 &  17.9$\times$ &     0.007 &  29.6$\times$ \\
          7) & paw\_P\_ai &     2.6 &    0.08 &  35.2$\times$ &     0.03 &  89.5$\times$ \\
          8) & paw\_P\_ap &     6.6 &    0.2 &  27.7$\times$ &     0.1 &  43.2$\times$ \\
            9) & paw\_add &    42.8 &    2.3 &  18.4$\times$ &     1.5 &  27.8$\times$ \\
     10) & optical\_limit &    50.7 &    6.2 &   8.1$\times$ &     5.6 &   9.1$\times$ \\
               11) & zherk &    89.6 &    6.9 &  12.9$\times$ &     5.4 &  16.6$\times$ \\
\hline
& Total, $\ve q \rightarrow 0$ &  188.5 &    20.9 &   9.0$\times$ &       18.1 &  10.4$\times$ \\ 
& Total, $\ve q \neq 0$ &  123.0 &    14.4 &   8.5$\times$ &       12.2 &  10.0$\times$ \\ 
\hline \hline
\end{tabular}
\end{center}
\end{table}


Table \ref{Tb:multix} (row 11) shows the timing and speed up 
of \textsc{zherk} compared to \textsc{zher} with different numbers of
$u$ ($Nu$), for $n(\ve G)=1587$. The speed up  is $12.9\times$ with
$Nu=50$ and $16.6\times$ with $Nu=250$. This improvement means the
other steps such as  9 and 10 are now the performance bottlenecks.
Thus we applied a similar idea to all the other steps. For simplicity,
we keep the size of $u$ uniform across the code.

As shown in Table \ref{Tb:multix}, the largest speed up comes from the
PAW portion.  One subtlety with this code is that the number of
projector functions is different for the different atomic species. In
the CPU code, one has to loop over atoms and perform the operations
sequentially for each atom, while in the GPU code the loop over
atoms, as well as the loop over projector functions (and bands when
using the ``multi-$u$'' approach) can be eliminated by using thread
parallelization. Each thread corresponds to an unique atom, projector
function and band index. Since the operations on each atom are still
linear algebra, the above algorithm is similar to batched
\textsc{cublas} function calls, although the sizes of the matrices are
different within the batch. Since such a ``flexible" \textsc{cublas}
batch is not yet available, we implemented our own customized
\textsc{cuda} kernel and achieved a significant improvement.  In
addition to the PAW and \textsc{zherk} portions, step 10 also has a non-trivial
speed up of $9.1\times$, which results from a combination of
batched \textsc{cufft} and our own kernels based on the ``multi-$u$''
approach.

The advantages of grouping small amount of independent data include
that the number of kernel launches is reduced, reducing the effect of
kernel launch overhead.  Also $cudaMemcpy$ is executed with larger
amount of data per copy.  This transforms the memory bound problem
into a computation/latency bound problem.

\subsection{Enhancing the CPU implementation}
The above techniques for transforming memory bound problems into
computational/latency bound problems can also be applied to the CPU
implementation.  In particular, a transformation from 
\textsc{zher} to \textsc{zherk} is straightforward. However, the
thread parallelization over atoms in the PAW part is not possible for
the CPU implementation.  For the optical limit portion, the thread
parallelization over bands and batched FFTW calls are non-trivial and
are expected not to gain much performance in the CPU implementation.
As a result, we only  implemented \textsc{zherk} on the CPU.
Given the same vector/matrix  used in Table \ref{Tb:nmultix1} and
\ref{Tb:multix}, the \textsc{zherk} versus \textsc{zher}
speed up is $6.6\times$ and $7.1\times$ for $Nu=50$ and 250, respectively,

\subsection{Final GPU/CPU Performance Improvement}

After enhancing both the CPU and GPU implementation using the
``multi-$u$'' approach, we summarize the 8-GPUs/8-CPUs speed up in
Table \ref{Tb:final_speedup} as a function of $Nu$. The
$Nu=1$ column is the same as the last column in Table
\ref{Tb:nmultix1}, which corresponds to the ``direct GPU porting"
without employing the ``multi-$u$'' approach. As $Nu$
increases, steps 3 - 5 have only slight changes and fluctuations. The
major speed up as a function of $Nu$ comes from the PAW part, which
enables simultaneous thread parallelization over atoms, projector
functions and bands. The speed up of \textsc{cublas} vs.
\textsc{blas} \textsc{zherk} increases from $13.3\times$ up to
$26.9\times$ at $Nu=250$. The matrix we used here (dimension of
1587$\times$1587) is still not large enough to achieve the peak
performance, which is around $36\times$. The final
speed up (in asynchronous mode) is $30.6\times$/$39.6\times$ for
optical/non-optical limit for the test system of N$_2$/Ru(0001)
surface.

\begin{table}[t]
\caption{\label{Tb:final_speedup} Final 8-GPUs vs 8-CPUs speed up after enhancing both the CPU and GPU implementation using the ``multi-$u$'' approach, with $Nu=$1, 2, 5, 10, 50, 150 and 250.  The same steps and test system are used as in Table \ref{Tb:nmultix1}. 
}
\vspace{0.2cm}
\begin{center}
\begin{tabular}{rlccccccccc}
\hline\hline
 No. & Function  & $Nu=1$   & 2 & 5 & 10 & 50 & 150 & 250 \\      
\hline   
            3) & get\_wfs &   11.6$\times$ &  15.3$\times$ &  17.6$\times$ &  15.2$\times$ &  14.2$\times$ &  15.4$\times$ &  16.0$\times$ \\
      4) & transform\_wfs &   12.7$\times$ &  14.8$\times$ &  18.8$\times$ &  13.4$\times$ &  14.3$\times$ &  14.8$\times$ &  15.4$\times$ \\
                 5) & fft &   11.1$\times$ &  12.6$\times$ &  14.8$\times$ &  12.5$\times$ &  12.1$\times$ &  12.4$\times$ &  12.3$\times$ \\
                6) & mapG &    2.2$\times$ &   4.4$\times$ &  10.9$\times$ &  14.0$\times$ &  33.8$\times$ &  47.5$\times$ &  53.4$\times$ \\
          7) & paw\_P\_ai &    2.8$\times$ &   5.0$\times$ &  12.1$\times$ &  23.4$\times$ &  94.0$\times$ & 195.9$\times$ & 242.2$\times$ \\
          8) & paw\_P\_ap &    1.5$\times$ &   2.6$\times$ &   7.0$\times$ &  10.5$\times$ &  36.2$\times$ &  51.5$\times$ &  56.7$\times$ \\
            9) & paw\_add &    5.6$\times$ &  13.0$\times$ &  30.7$\times$ &  48.6$\times$ &  82.1$\times$ & 114.8$\times$ & 136.2$\times$ \\
     10) & optical\_limit &    5.3$\times$ &   4.7$\times$ &   9.5$\times$ &  11.2$\times$ &  17.8$\times$ &  19.9$\times$ &  19.2$\times$ \\
               11) & zherk &   13.3$\times$ &  13.0$\times$ &  13.2$\times$ &  11.8$\times$ &  22.9$\times$ &  26.2$\times$ &  26.9$\times$ \\
\hline
& Total, $\ve q \rightarrow 0$ &   9.6$\times$ &   11.2$\times$ &   15.8$\times$ &   16.8$\times$ &   26.4$\times$ &   29.9$\times$ &   30.6$\times$ \\ 
& Total, $\ve q \neq 0$         &  12.5$\times$ &   13.6$\times$ &   18.8$\times$ &   20.9$\times$ &   31.8$\times$ &   37.8$\times$ &   39.6$\times$ \\ 
\hline \hline
\end{tabular}
\end{center}
\end{table}

\section{Performance across different systems}\label{sec:performance}
The goal of the GPU port is not merely to improve performance, but to
be able to address scientific problems with RPA. In this section we
examine the performance of the GPU implementation across three types
of systems: molecules, bulk solids and molecules adsorbed on surfaces.

\begin{table}[t]
  \caption{\label{Tb:systems} The 8-GPUs/8-CPUs speed up (column ``Speed up", for $\ve q \neq 0$) as well as the time required to complete the entire RPA calculation with a response function cutoff of 150 eV (column $t_{gpu}$) using 8-GPUs for different systems (column ``System"). For each system, the phase (column ``Phase"), the number of atoms (column $N_a$) and the number of electrons (column $N_e$) in the unit cell, whether it is spin polarized (column ``Spin") and the number of $k$-points sampled for the BZ are specified. ``sec" and ``h" stand for seconds and hours, respectively. 
  }

\vspace{0.2cm}
\begin{center}
\begin{tabular}{lcccccccc}
\hline\hline
 System & Phase & $N_a$ & $N_e$ & Spin & $k$-points & Improvement & $t_{gpu}$\\      
\hline   
O$_2$ & gas & 2 & 12 & True & 1 & 11.3x & 41 sec \\ 
Li$_2$O             & bulk & 3   & 8   & False & $4\times 4\times 4$   &  10.5x & 63 sec \\ 
MoO$_{3}$ & bulk & 16 & 96 & False & $4\times 2\times 4$   &  35.3x   & 1.0 h \\ 
N$_2$/Ru(0001) & surface & 14 & 202 & False & $4\times 4\times 1$ & 36.1x & 1.4 h\\ 
CO/Ni(111) & surface & 22 & 210 & True & $4\times 4\times 1$ & 37.0x   & 5.5 h\\ 
\hline \hline
\end{tabular}
\end{center}
\end{table}

Table \ref{Tb:systems} summarizes the selected systems, their
performance improvement and the time required to complete the RPA
calculation with a response function energy cutoff 150 eV.  This
cutoff is not high enough for a fully-converged result (250 eV is the
minimum energy cut off to get converged results up to 50 meV) but this
does not affect the conclusions here, since the speed up is in
principle more favorable with larger energy cutoff, which corresponds
to a larger number of plane waves and thus larger matrices.  Also, the
results are useful for comparison between similar systems or
extrapolation of the results to different numbers of $k$-points used in
the simulation. The number of bands used is equal to the number of
plane waves for all the RPA calculations. Since a full CPU calculation
is time consuming and unnecessary, the speed up presented in the
table is obtained by performing some of the identical loops in the response
function summation, while the time $t_{gpu}$ is obtained by completing
the entire RPA calculation on 8-GPUs. The multi-$u$ approach with
$Nu=100$ is used for both the CPU and GPU calculations.

O$_2$ is selected as a representative molecule.  Although the number
of plane waves increases linearly with the volume of the cell, it is
reported that the RPA correlation energy converges rather fast with
respect to the vacuum size used in the cell \cite{Kresse_B08}.  Using
a (7\AA, 7\AA, 8.3\AA) simulation cell, the RPA calculation of O$_2$
can be finished in 41 seconds on 8-GPUs. Similar timing applies to
other small molecules such as N$_2$ and CO.


For bulk systems, a simple metal oxide Li$_2$O and a transition metal
oxide MoO$_3$ are selected. The unit cell for MoO$_3$ consists of 16
atoms and is larger than that for Li$_2$O. The speed up increases from
$10.5\times$ to $35.3\times$ due to the larger unit cell (because a
larger number of plane waves are used). The total GPU time changes
dramatically from 63 seconds for Li$_2$O to 1 hour for MoO$_3$ due to the theoretical $O(N^4)$ scaling for RPA calculations.
The speed up and GPU timing on these bulk metal oxides are encouraging
considering that we have spent over 1.5 million computing hours for
calculating the formation energies of 23 metal oxides
\cite{Jun_oxide}.

For surface systems, we selected two representative examples: N$_2$
adsorbed on Ru(0001) and CO on Ni(111) surfaces. The Ru(0001) surface
is modeled with 4 layers having a $\sqrt{3} \times \sqrt{3}$ unit cell
and the Ni(111) surface with 5 layers and a $2\times 2$ unit cell. The
latter unit cell and number of layers used are reported to converge
the DFT chemisorption energies well \cite{Jess_beef}.  The vacuum
region is set to be 15 \AA~for both surfaces. Semi-core (4$s$ and 4$p$)
states are included in the Ru(0001) PAW potential. As a result, the
number of electrons is similar for the Ru(0001) and Ni(111) surfaces,
although the latter has a larger number of atoms included in the unit
cell. A spin polarized calculation is employed for the Ni(111) case.
The performance improvements for both systems are slightly better than
for bulk MoO$_3$. It suggests that the speed up is approaching,
although not quite achieving, the maximum possible, without
fine-tuning of $Nu$ on a per-system basis. The final GPU time is 1.4
and 5.5 hour for N$_2$/Ru(0001) and CO/Ni(111), respectively.  The longer
time for CO/Ni(111) is because it is a spin polarized calculation with
a larger $2\times 2$ unit cell, which results in an increased number
of bands included in the response function summation.

\section{Conclusion and Outlook}\label{sec:conclusion}
We have ported the non-interacting density response function onto the
GPU architecture. By grouping independent charge density matrices, we
transformed the problem from being memory bound into being
computation/latency bound.  We call this the ``multi-$u$'' approach,
where the number of $u$ is the number of independent vectors/matrices
grouped together. The number of $u$ is flexible and constrained only
by available memory. We enhanced both the CPU and GPU implementations
with the ``multi-$u$'' approach.  The RPA calculations remain on the GPU
(no ``thunking'').  The size of the code is roughly 6000 lines of
\textsc{python} and 1000 lines of \textsc{c/cuda} (many \textsc{gpaw}
functions are re-used and not counted here)\footnote{The RPA GPU implementation is available for download at the GPAW svn repository \\ \url{https://trac.fysik.dtu.dk/projects/gpaw/browser/branches/rpa-gpu-expt} }.  The RPA correlation
energy calculation performance improvement (8-GPUs vs 8-CPUs) is
around $10\times$ for very small systems, and $40\times$ for standard
bulk systems and surfaces. With this improvement, an RPA calculation of CO
adsorbed on Ni(111) surface using 5 layers and $2\times 2$ unit cell,
sampled with 16 $k$-points, can be finished in 5.5 hours using 8 GPUs.
Such a speed makes it promising to employ non-self-consistent RPA for routine surface
chemistry simulations, although it should be noted that the $O(N^4)$
scaling for RPA calculations have not changed by porting to GPU.
Furthermore, since the non-interacting response function is one of the
most important and time consuming ingredients for many beyond-DFT
calculations such as TDDFT, GW, Bethe-Salpeter\cite{Jun_BSE}, we
expect similar performance improvements in these
beyond-DFT calculations using the ``multi-$u$'' approach.

\section{Acknowledgements}
J. Y. thanks Phillippe Vandermersch for suggesting the use of
\textsc{cublas} \textsc{zherk} instead of \textsc{zher}, Lung Sheng
Chien for providing help with \textsc{cublas} and validation of the
CPU and GPU timing, and Marcin Du{\l}ak and Jens K. N{\o}rskov for
commenting on the manuscript.  The authors acknowledge hardware
donations from Nvidia Corp., and support by the Department of Energy,
Office of Basic Energy Sciences, under contract DE-AC02-76SF00515.

\bibliographystyle{model1a-num-names}

\end{document}